\documentclass[epj]{webofc}
\usepackage[varg]{txfonts}   
\usepackage{lineno}
\begin{document}
%
%
\title{Seasonal variations of the rate of  multiple-muons   in the Gran Sasso underground laboratory}
%
%

\author{F.~Ronga\inst{1}\fnsep\thanks{\email{francesco.ronga@lnf.infn.it
    }}        }

\institute{Istituto Nazionale di Fisica Nucleare, Laboratori Nazionali di Frascati, via E.Fermi~40 - 00044 Frascati, Italy         }

\abstract{%
  It is  well known that the rate of cosmic ray muons depends on the atmospheric temperature, and that for events with a single muon  the peak of the rate is in summer, in underground laboratories in the northern hemisphere.
In 2015 the MINOS  experiment,  in USA,  found that, for small distances between  the multiple-muons, the rate  of multiple-muons peaks  in the winter and that the amplitude of the modulation is smaller than  in the case of a single muon.  I have done a re-analysis of  data of the past MACRO experiment. The result is that under Gran Sasso the rate of multiple-muons at small distances peaks in the summer. This difference with MINOS could be explained by differences in the atmospheric temperature due to latitude.
This results could be of  interest for dark matter experiments looking to  dark matter seasonal modulation due to the Earth's motion.

}
\maketitle
\section{Introduction}
\label{intro}

Underground muons originate primarily from the decay of mesons produced in high energy interactions
between primary cosmic ray particles and atmospheric nuclei \cite{Ga1990}. 
Fluctuations in atmospheric temperature lead to variations in the muon rate observed at ground level and underground.

While the temperature of the troposphere varies considerably within the day, the temperature of the stratosphere remains nearly constant during the day, usually changing on the time scale of seasons.
An increase in temperature of the stratosphere causes a decrease in density.
This reduces the chance of meson interactions, resulting in a greater number of mesons that  decay and produce muons and increases the muon rate observed by several experiments located deep underground.
For the a recent summary of the data see  \cite{Agostini:2016gda}. 
The majority of muons detected in an underground detector are produced in the decay of pions.
All the underground experiments until 2015  analyzed the muon rate of events with a single track, 
or with specific cuts to select those events as for example in \cite{Ambrosio:1997tc}  or simply
because in small detectors the rate is largely dominated by the single muons. 


In 2015 the MINOS experiment published the seasonal modulation of the multiple-muons events \cite{Adamson:2015qua}.
We expect different seasonal oscillation between multi muon events and single muon events. One difference is due to the fact that multiple-muons are produced by primary cosmic rays with an energy higher than the one needed for single muons. Another difference is due 
to the different cosmic ray primary, multiple-muons production is preferred by heavy primaries. Heavy  primaries have a cross section 
 bigger than proton and therefore interact higher in the atmosphere. Another effect is geometrical: large size multiple-muons 
are produced  at a bigger heigth than single muons or multiple muons of a small size. Of course this kind of geometrical effect  depends
also by the size of the detector. 

MINOS reports results of two detector: a far detector (FD) with an overburden of 225 m.w,e, and a near detector (ND) at 2100 m.w.e.
MINOS has observed a multiple-muon  phase inconsistent with the summer maximum observed in the ND and the  FD single-muon data. Data collected by the MINOS FD were used to show that there is a transition from a summer maximum in multiple-muon events with a large track separation to a winter maximum in multiple-muon events with a small track separation. This transition occurs at track separations of about 5-8 m. The amplitude of the oscillation was also dependent on the tracks separation.
 

The dependence of the muon rate variations on the atmospheric temperature at the first order can be  expressed  as  \cite{Ambrosio:1997tc} :





\begin{equation}
\frac {\Delta I_\mu} {I^{0}_{\mu}}= \alpha_{T} \frac {\Delta_{Teff}}{<T_{eff}>} 
\label{Teff}
\end{equation}

where $\alpha_{T}$ is a constant and  $T_{eff}$ is an $effective$ temperature. The effective temperature can be computed dividing the atmosphere in layers for which  the temperature measurements exist. The effective temperature is a weighted mean of the layer temperatures. To compute $T_{eff}$  I have used    the formula   given in the MACRO paper \cite{Ambrosio:1997tc}. The  $T_{eff}$  used by MINOS \cite{Adamson:2015qua} looks different, but numerically gives similar results to the one used by MACRO. For the atmospheric temperature I have used the temperature measurements  at 37 atmospheric pressures provided by the ECWMF, the European Centre for Medium-Range Weather Forecasts \cite{ecwmf}.

MACRO  \cite{Ambrosio:2002mb} has been the largest acceptance  cosmic ray detector located in the Gran Sasso underground laboratory
in Italy.  
The large acceptance of MACRO allowed a large number statistics for multiple-muons.
 MACRO ended data taking in December 2000. 
The data presented on the single muon seasonal variation in the paper \cite{Ambrosio:1997tc} were collected during the MACRO construction in the period December 1992-December 1994. The data presented in this paper have been collected in the period December 1995 -December 2000 with the full MACRO, corresponding to about $34.5 \times 10^{6}$ single muons and and $2.6 \times 10^{6}$
events having at least 2 tracks. 





\begin{figure}
\centering
\includegraphics[width=14.5cm,height=7.5cm]{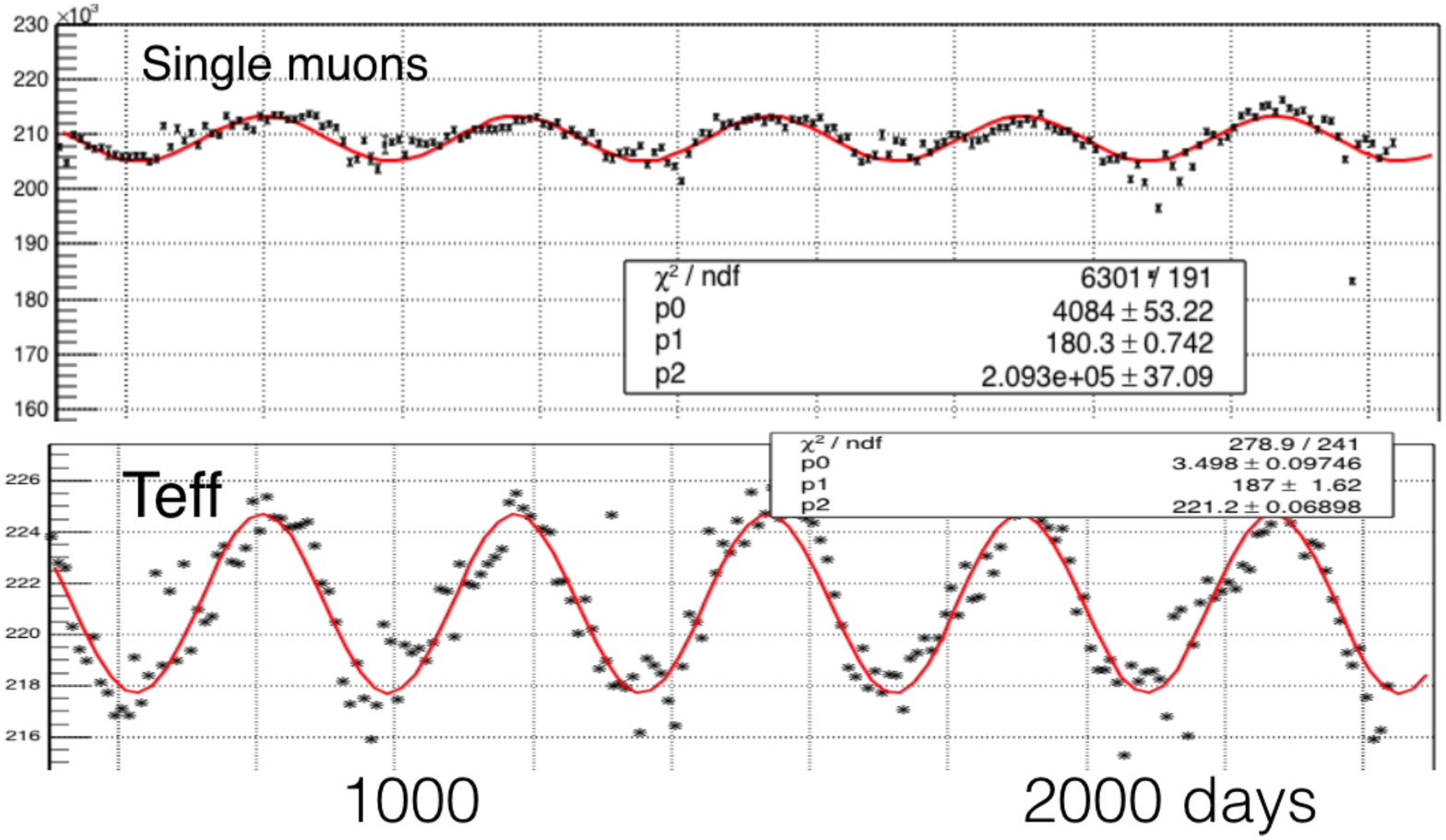}
\vskip-1.2cm
\caption{Single muons every 10 days. The x axis shows the day number starting from April 20th 1994. On the top are the single  muons every 10 days. The bottom plot shows the effective temperature $T_{eff}$ in Kelvin. The fit is done with a sinusoid +  a  constant (p2). The parameters p0/p2 = 1.9\% give the  oscillation amplitude, p1 gives the day of the  peak, starting from day 0 =January 1th. p1 is 180.3 for single muons,  The periodicity is fixed at 365.2 days}
\label{fig1}       
\vskip-0.5cm
\end{figure}

\section{The data selection}
 The tracking package used in this analysis requires at least 4 horizontal streamer chamber planes
 or at least  2 horizontal streamer planes and 2 vertical streamer planes \cite{Ambrosio:2002mb}.
In this analysis I have used only one view: the streamer tube "wires view"; this in order to avoid the
problems due to the possible wrong association of the two different views to define tracks in
multiple-muons events.
\begin{figure}[b]
\vskip-0.2cm
\centering
\includegraphics[width=14.5cm,height=9.0cm]{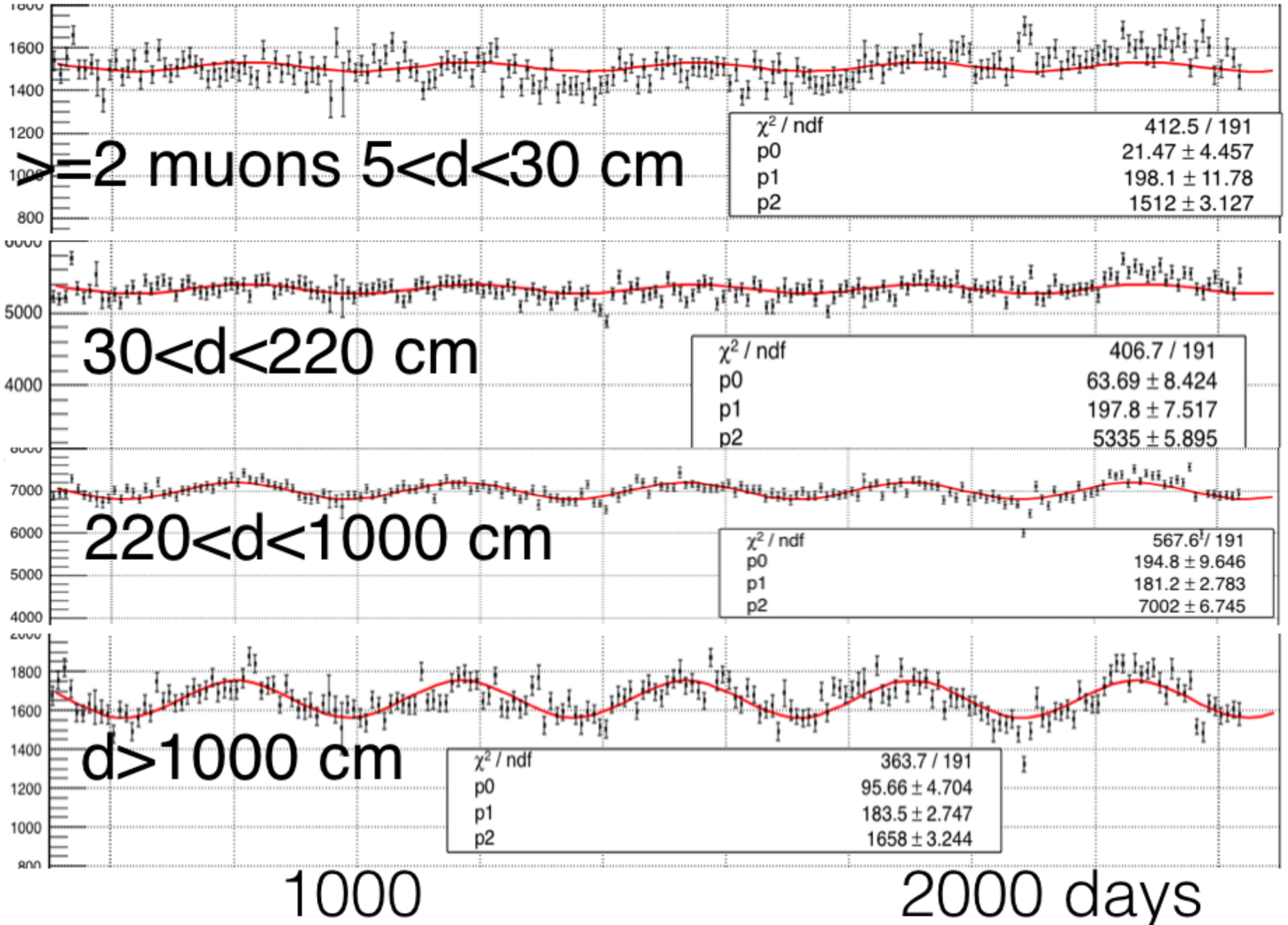}
\vskip-0.3cm
\caption{Multiple-muons every 10 days. The x axis shows the day number starting from April 20th 1994. Starting from the top:   muons every 10  events with at least 2 tracks  and with average distance between 5 and 30 cm, 30 and 220 cm, 220 and 1000 cm and average distance bigger than 1000 cm.  The day of the peak of the rate (parameter  p1)  varies from 181 to 198 (July 1th- July 18th.)}
\label{fig2}       
\end{figure}

This analysis is looking for effect at the $1\%$ level. This means that is important to select runs having good efficiencies.
MACRO was divided in 6 "supermodules" and the data acquisition was done with 3 separate data acquisitions,
each collecting data from two supermodules. Sometimes a couple of supermodules was removed	
from the acquisition for the maintenance of the detector. Since the rate of single muons  with the full MACRO is
about 860 events/hour it is possible to select the runs with full MACRO applying a cut on the rate of the single
muons. So I have analyzed only the runs with rate of the single muons between 710 and  1010 events/hour  corresponding to a cut of more than 5 $\sigma$, a factor 9 larger than the $1.9$\%  single muon variation due to the seasonal modulation.

 I have not tried more sophisticated selection based on the logbook of the detector performance because  of technical problems. I recall that the MACRO data analysis was designed around 1990 and it was based  on alfaVAX Digital Equipment computers with VMS operatinfg system. A fraction of the analysis code, but not all, was ported to UNIX before the end of the experiment. So in this analysis it was difficult to access to all the information on the data quality and to the detailed data base of the detector efficiencies. I have used only the information contained on data summary files containing track and hit informations. 

However it is important to note the  requirement on the very small number of hits to define a track (4 over a maximum possible of 14 for the horizontal chambers, or 2 horizontal and 2 vertical). Since the typical streamer tube efficiency is of the order of $97\%$,
reasonable variations of this efficiency doesn't change very much the tracking efficiencies.
MACRO started to take data with the full detector around April 20 1994, but in this analysis I have used only data starting from December 1995,  to select data collected in  stable conditions.

\section{Results}

\label{results}
The number of the single muon events every 10 days is shown at the top of Fig.~\ref{fig1}. The x axis of the  plot is the day number (day 0 is  April 20 1994). The fit is done with a sinusoid + a constant value (parameter p2). The periodicity is fixed at the  value of 365.2 days. The fit parameters  ratio p0/p2 gives the fractional oscillation amplitude, p1 gives the day of the  peak. p1 is 180.3 for single muons (Jun 30th-July 1th).
From the daily correlation of the single muon rate with $T_{eff}$  of Eq ~\ref{Teff} is obtained $\alpha_{T}=1.03\pm0.01(statistical)$ higher than the  value $\alpha_{T}=0.83\pm0.13$ reported in  the MACRO seasonal modulation paper\cite{Ambrosio:2002mb},to be compared the theoretical value $\alpha_{T}=\sim0.92$  \cite{Agostini:2016gda}.

In case of multiple-muons the distance between couples of tracks has been evaluated using only the wire view, and the average value of the distances has been evaluated for each event.  Fig.~\ref{fig2} shows  the rate of the multiple muons with different cuts on the average distance:  distance between 5 and 30 cm, 30 and 220 cm, 220 and 1000 cm and average distance bigger than 1000 cm.  In the case of the multiple muons the peak of the rate  varies  from day 181, for distances bigger than 1000 cm, to day 198, for distances between 5 and 30 cm. The peak of the rate of multiple muons is delayed of a few days respect to the one of single muons, but is far from the winter months as observed in MINOS. Also the amplitude varies from a minimum of 0.4\% to a maximum of 5.8\%  for distances bigger than 1000 cm.

\begin{figure}
\begin{center}
 \includegraphics[height=85mm,width=120mm]{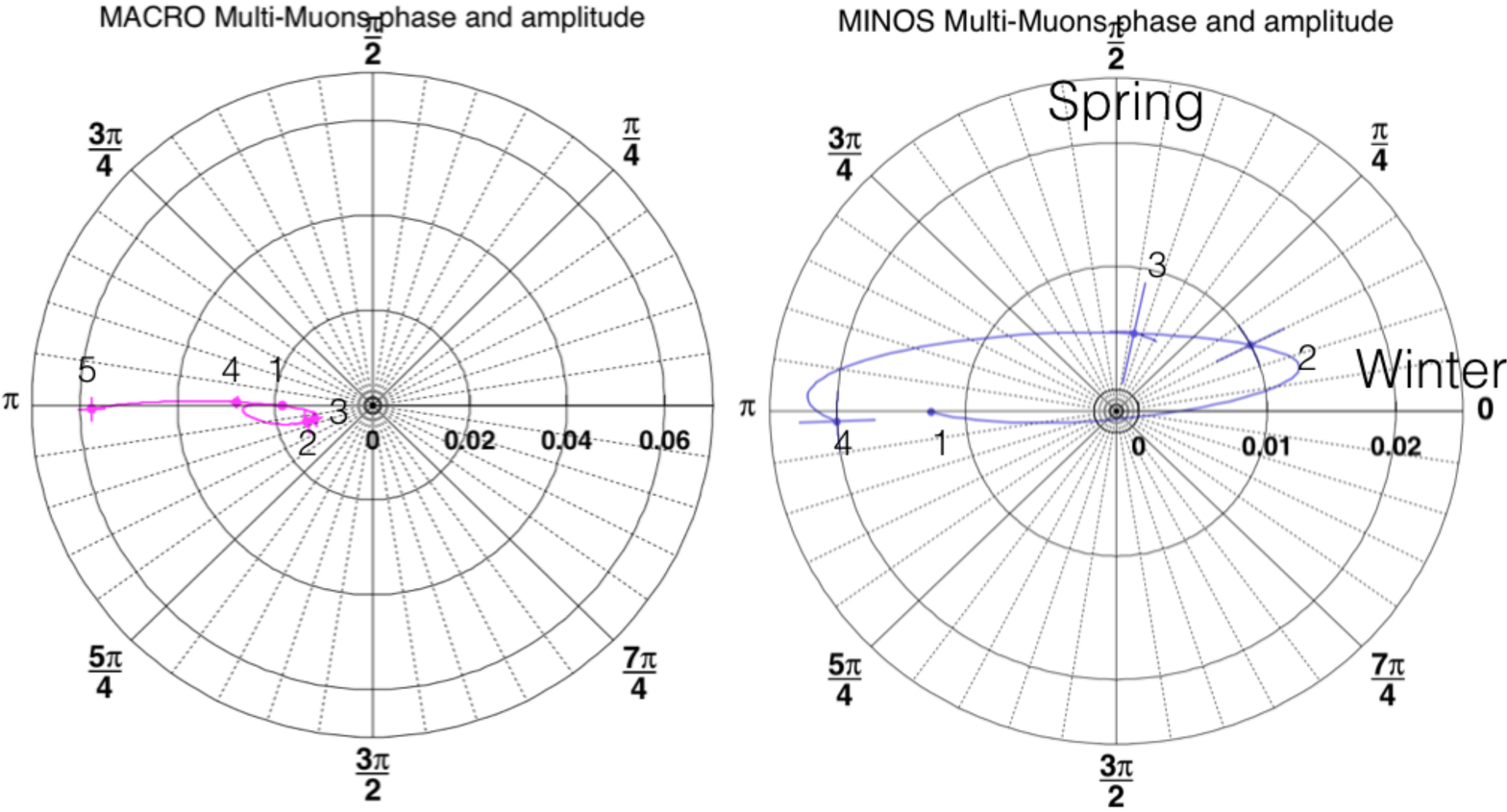} 
\end{center}
\vskip-1.8cm
\caption { The amplitude and the phase of the first harmonic of the multiple muon rate in MACRO and MINOS. Phase 0 is near January 1th. Phase $\pi$ is near July 1th. The line is only to guide the eye. The points labeled as 1 show  the value for the single muons. The points after 1 are  for multiple-muons. They are  in increasing values of the separation between muons. In MACRO the points 2-5 corresponds to cuts   in the average muon distance 5-30, 30-220,220-1000 cm , $\ge$1000.  In MINOS the cut is in the minimum distance and the points 2-4 corresponds to cuts 60-450, 450-800,$\ge$800 cm. In MINOS there are big changes in the phase, but not in MACRO.  In  both experiments, the last point (cut at large distance) has  practically  the same phases of the one of single muons, while the amplitude is larger. Note that the radial scale of the two plots is different.}
 \label{fig3}
\end{figure}

The results are seen in a better way in the polar graph of figure Fig.~\ref{fig3}. On the left are the MACRO results and on the right are the MINOS results. The MACRO and MINOS data looks quite similar: in both there is a small amplitude at short distances and the oscillation amplitude is larger then the one of single muons at large distance (this is expected if every muon of multi-muons events is produced as "single" muon). In both experiments there are changes of the peak position correlated with the amplitude, but in MACRO the variations of the peaks are much smaller that in MINOS, while the variations of the amplitude are similar. 

A comparison of the  ECWMF atmospheric temperature in 37 layers shows different seasonal variations in the two experimental sites. In fact in  MINOS  there are atmospheric layers with a peak temperature in the winter season, while in MACRO the peak temperature is always in the summer months. This differences occurs at a height of about 13 km, where atmospheric jet streams are important. This difference is probably due to the different latitudes ( $42^{0}$  for Gran Sasso and  $48^{0}$ for MINOS FD ).   

\section{Conclusion}
The peak of the rate of the multiple muons under Gran Sasso with different cuts on the distance, are  in July, delayed respect to the one of single muons, while the expected peak of the dark matter signal is expected around June 2;  therefore multiple-muons originated backgrounds should not be a problem for the DAMA \cite{Bernabei:2013xsa} dark matter experiment. This result is different in the MINOS FD site, the difference could be due to the different depths and to the different latitudes.

The author thanks A. Longhin,  for giving me the ECWMF temperatures,  A. Paoloni and A.Marini for useful suggestions and discussions, and  all the MACRO past collaborators,  listed in ref \cite{Ambrosio:2002mb}, contributing to the success of this experiment. This work also shows the importance to save  data and the software of past experiments.


%
%

\end{document}